\documentclass{acta}

\usepackage{amsmath}
\usepackage{amssymb}
\usepackage{rotating}
\usepackage[hidelinks]{hyperref} 
\usepackage{bm}

\usepackage{xcolor}

\newcommand{\dummy}[1]{}

\begin{document}

\begin{Titlepage}

\Title{OGLE Pre-discovery Observations of the Dwarf Planet Candidate 2026 ED138}

\Author{Przemek Mr\'oz$^1$, Andrzej Udalski$^1$, Igor Soszy\'nski$^1$, \\ Micha\l{}~K. Szyma\'nski$^1$, Pawe\l{} Pietrukowicz$^1$, Rados\l{}aw Poleski$^1$, \\ Jan Skowron$^1$, Krzysztof Ulaczyk$^{2,1}$, Mariusz Gromadzki$^1$, \\Krzysztof Rybicki$^1$, Patryk Iwanek$^1$, Marcin Wrona$^{3,1}$, Mateusz J. Mr\'oz$^1$}
{$^1$Astronomical Observatory, University of Warsaw, Al. Ujazdowskie 4, 00-478 Warszawa, Poland\\
$^2$Department of Physics, University of Warwick, Coventry CV4 7AL, UK\\
$^3$Department of Astrophysics and Planetary Science, Villanova University, 800 East Lancaster Avenue, Villanova, PA 19085, USA}

\Received{MM DD, YYYY}

\end{Titlepage}

\vspace*{-10pt}
\Abstract{
The newly discovered trans-Neptunian object, 2026~ED138, is a strong dwarf planet candidate detected by the Vera C. Rubin Observatory. Despite its high intrinsic brightness, it evaded detection by previous sky surveys as it hid near crowded Galactic disk fields. Here, we present an analysis of precovery observations of 2026~ED138 in archival data from the Optical Gravitational Lensing Experiment (OGLE). We identify 102 reliable detections of the object as it moved across the dense regions of the Galactic plane and bulge between 2011 and 2024. We provide precise astrometric and photometric measurements of the object. Modeling its phase curve yields absolute magnitudes of $H_I = 2.224 \pm 0.034$\,mag and $H_V = 3.154 \pm 0.035$\,mag. Period analysis of the phase-angle-corrected light curve reveals a candidate rotation period of 1.18077\,days and its 1-day alias 0.54066\,days, although a low signal-to-noise ratio renders these periodicities tentative.}
{Trans-Neptunian objects (1705), Kuiper belt (893), Dwarf planets (419) }

\section{Introduction}

The newly detected trans-Neptunian object (TNO), 2026~ED138, is one of the intrinsically brightest known objects in its class, making it a strong dwarf planet candidate. It was discovered in data from the Vera C. Rubin Observatory by \citet{holman2026}, who derived an absolute magnitude of the object of $H_V=3.19 \pm 0.04$\,mag, corresponding to an estimated diameter of 612--1530\,km for geometric albedos between 0.25 and 0.04. By combining Rubin observations with archival DECam data spanning 2016 to 2023, \citet{holman2026} determined a semi-major axis of $a \approx 43.09$\,au, an orbital eccentricity of $e \approx 0.139$, and an inclination of $i \approx 24.4$\,deg, placing the object near the edge of the 12:7 mean-motion resonance with Neptune.

Shortly after the discovery, \citet{han2026} reported detections of the object in full-frame images from the Transiting Exoplanet Survey Satellite (TESS) taken between 2019 and 2026. \citet{murtagh2026} analyzed publicly available archival images from various sky surveys, identifying 2026~ED138 on 13 photographic plates (taken between 1937 and 2000) and in several CCD images from surveys such as VISTA, SkyMapper, and ATLAS (from 2010 on). Incorporating these historical observations significantly refined the orbital solution for the object. \citet{murtagh2026} argued that about half of this improvement can be tracked to precise astrometry from VISTA and SkyMapper, highlighting the usefulness of archival CCD observations.

One reason why 2026~ED138 escaped detection in previous TNO searches is the fact that it lies deep in the southern hemisphere near the Galactic plane, a region typically avoided by traditional TNO surveys. An exception to this rule was the OGLE-Carnegie Kuiper Belt Survey \citep[OCKS;][]{sheppard2011}, which was conducted as part of the fourth phase of the Optical Gravitational Lensing Experiment (OGLE-IV; \citealt{udalski2015}). The OCKS survey was carried out between 2010 March and 2011 January using the 1.3-m Warsaw Telescope equipped with a wide-field mosaic CCD camera.

The OCKS survey led to the discovery of several large TNOs, including 471143 Dziewanna (a.k.a. 2010 EK139). With an absolute magnitude of $H_V \approx 3.8$\,mag, this object was initially estimated to have a diameter as large as 1000\,km. However, subsequent thermal observations with the \textit{Herschel} satellite indicated a much smaller diameter of $470^{+35}_{-10}$\,km \citep{pal2012} or $433^{+63}_{-64}$\,km \citep{lellouch2013}, while a stellar occultation observed on May 17, 2019 indicated a minimum major-axis length of 507\,km \citep{burzynski2024}.


The OCKS survey observed regions within 20\,deg of the ecliptic plane. Curiously, at that time, 2026~ED138 was located at an ecliptic latitude of approximately $-23$\,deg, thereby evading detection in 2010. Nevertheless, over the past 16 years, 2026~ED138 remained close to the Galactic plane and Galactic bulge fields that were regularly observed as part of the OGLE-IV survey. This article presents an analysis of these pre-discovery OGLE observations of 2026~ED138.

\section{Data}

All data reported in this paper were collected as part of the OGLE-IV survey \citep{udalski2015}. The survey operates the 1.3-m Warsaw Telescope located at Las Campanas Observatory\footnote{Minor Planet Center observatory code 304}, Chile, which is equipped with a mosaic CCD camera providing a field of view of 1.4\,deg$^2$. The camera consists of thirty-two $2048 \times 4102$ detectors with a pixel scale of $0\zdot\arcs 2585$ per pixel. Most observations were taken through the $I$-band filter, which closely resembles that of the standard Cousins system. Approximately 10\% of images were taken through the $V$-band filter, whose transmission curve resembles that of the Johnson filter.

The survey observes a pre-defined grid of fields covering regions of the sky with the highest stellar density: the Galactic bulge, the Galactic plane, and the Magellanic Clouds. To identify images that might contain 2026~ED138, we first generated its ephemeris for the years 2010--2026 at 1-day intervals using the JPL Horizons System.\footnote{ https://ssd.jpl.nasa.gov/horizons/app.html} The ephemeris was generated based on the orbital solution that was available on November 2, 2026.

\begin{sidewaysfigure}
\includegraphics[width=\textwidth]{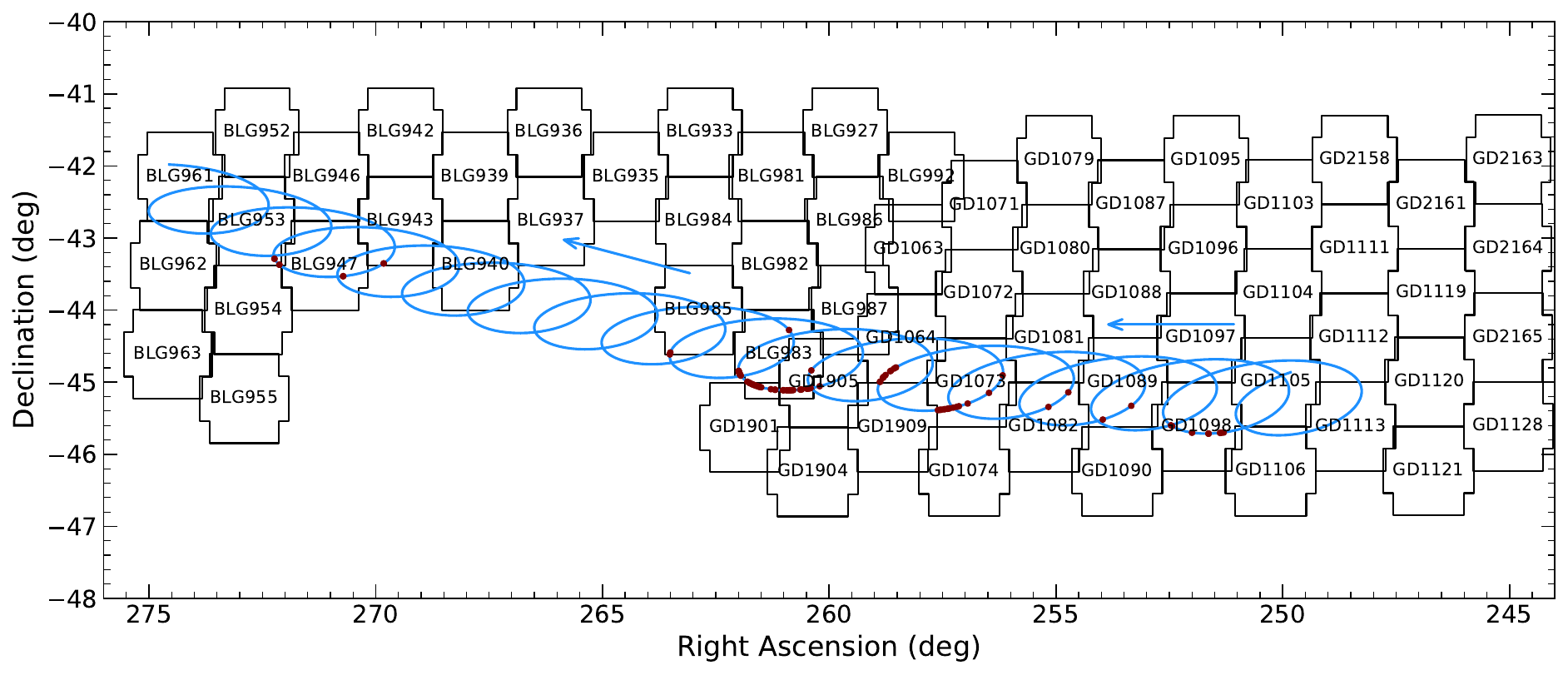}
\caption{Sky trajectory of 2026~ED138 from 2010 to 2026 (blue line). Black polygons mark the OGLE fields located along the path of the object. Pre-discovery OGLE observations are indicated by brown dots. Arrows indicate the direction of motion.}
\label{fig:map}
\end{sidewaysfigure}

Fig.~\ref{fig:map} shows the sky trajectory of 2026~ED138 along with the grid of nearby OGLE fields. The path of the object crosses the eastern Galactic disk fields (GD1064, GD1073, GD1081, GD1082, GD1089, GD1090, GD1097, GD1098, GD1105,\break  GD1106, GD1112, GD1113, GD1901, GD1905, GD1909) as well as the western fields of the shallow Galactic bulge survey (BLG940, BLG943, BLG946, BLG947, BLG953, BLG961, BLG962, BLG983, BLG985). We identified all available OGLE images of these fields, calculated the expected position of 2026~ED138 at the time of each observation, and checked whether the position fell within a given field. In total, we identified 128 images that were expected to contain 2026~ED138.

The OGLE photometric pipeline is based on the Difference Image Analysis (DIA) technique \citep{tomaney1996,alard1998} as implemented by \citet{wozniak2000}. A deep, nearly noiseless reference image is constructed for each field and filter by stacking several high-quality individual frames. This reference image is then subtracted from each science frame, producing a difference (or subtracted) image. As a result, all constant stars are subtracted out, leaving only objects that have changed their brightness relative to the reference image or changed their position.

\begin{figure}
\includegraphics[width=\textwidth]{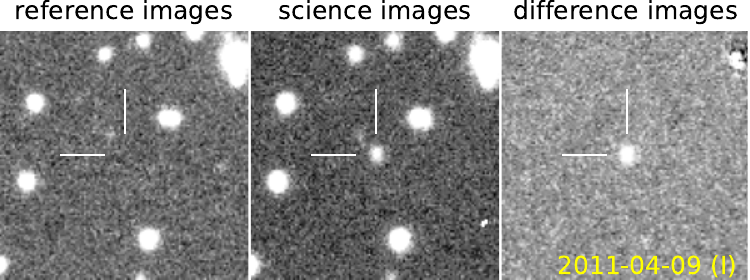}
\includegraphics[width=\textwidth]{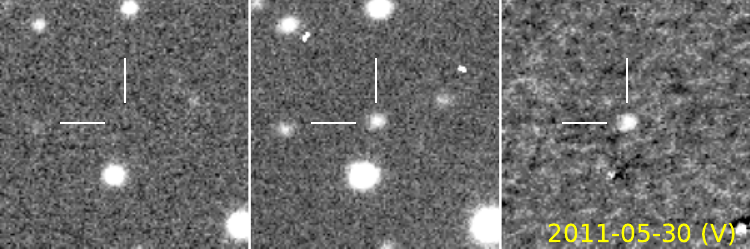}
\caption{Example pre-discovery images of 2026~ED138 taken on the night of April 9, 2011 ($I$ band; upper panel) and May 30, 2011 ($V$ band; lower panel) with the exposure time of 180\,s. The left column shows cutouts of the reference images, the middle column shows the science images, and the right column shows the difference images. The expected position of 2026~ED138 is marked by white tick marks. All cutouts are $30'' \times 30''$, North is up and East is to the left.}
\label{fig:fcharts}
\end{figure}

We examined the subtracted images for all frames that we identified in the previous step. We were able to detect the target near the position predicted by the JPL Horizons ephemeris in 112 of these images (for the remaining 16 images, the object was located slightly outside the footprint of the detector or in a gap between detectors of the CCD mosaic). The observed positions matched the predicted ephemeris very well, typically within 1--2 pixels (or $0\zdot\arcs25-0\zdot\arcs5$). Example images are shown in Fig.~\ref{fig:fcharts}, where the left column shows $30''\times30''$ cutouts of the reference image, the middle column shows the science images, and the right column shows the difference images. The expected position of the object is indicated by white tick marks.

In most cases, in particular in the shallow survey frames, the object was too faint to be detected by the standard OGLE DIA pipeline. Therefore, to measure its position and flux, we fitted a 2D Gaussian profile to a $20\times 20$\,pixel cutout around the expected position:
\begin{equation}
    F(x,y) = \frac{F_0}{2\pi \sigma^2}\exp\left(-\frac{(x-x_0)^2+(y-y_0)^2}{2\sigma^2}\right) + B,
\end{equation}
where $(x_0,y_0)$ is the position of the centroid, $\sigma$ is the standard deviation of the profile, $F_0$ is the flux of the object, and $B$ is the background flux. The best-fit parameters and their uncertainties were found using the Levenberg-Marquardt algorithm. We were able to obtain reliable flux and position measurements for 2026~ED138 in 102 images. For the remaining ten images, the object was either marginally detected or blended with a brighter background star that was not perfectly subtracted, biasing the measurements. 

All 102 measurements are reported in Table~\ref{tab:results}, which includes 91 epochs in the $I$ band and 11 in the $V$ band. The 2011 observations (six in $I$ and six in $V$) were obtained with exposure times of 180\,s. The remaining images were collected as part of the OGLE-IV Galaxy Variability Survey \citep{udalski2015} with much shorter exposure times (typically, 25\,s in $I$ and 30\,s in $V$).

\begin{table}
\caption{Positions and apparent magnitudes of 2026~ED138 measured from archival OGLE images.}
\label{tab:results}
\small
\begin{tabular}{rrrrrcrr}
\hline \hline
\multicolumn{1}{c}{JD (UTC)} & \multicolumn{1}{c}{RA} & \multicolumn{1}{c}{Decl.} & \multicolumn{1}{c}{$\sigma_{\rm RA}$} & \multicolumn{1}{c}{$\sigma_{\rm Decl.}$} & \multicolumn{1}{c}{Filter} & \multicolumn{1}{c}{$m$} & \multicolumn{1}{c}{$\sigma_m$}\\
& \multicolumn{1}{c}{(deg)} & \multicolumn{1}{c}{(deg)} & \multicolumn{1}{c}{($''$)} & \multicolumn{1}{c}{($''$)} & & \multicolumn{1}{c}{(mag)} & \multicolumn{1}{c}{(mag)} \\
\hline
2455660.69345 & 252.4650651 & --45.5974260 & 0.075 & 0.077 & I & 18.863 & 0.047 \\
2455662.70740 & 252.4381975 & --45.6084694 & 0.077 & 0.077 & V & 19.830 & 0.061 \\
2455686.60781 & 251.9990365 & --45.6993743 & 0.074 & 0.073 & I & 18.817 & 0.039 \\
2455686.62997 & 251.9985916 & --45.6994135 & 0.075 & 0.074 & V & 19.716 & 0.045 \\
2455701.70291 & 251.6366537 & --45.7131462 & 0.079 & 0.079 & I & 18.954 & 0.070 \\
2455701.72507 & 251.6361107 & --45.7131488 & 0.090 & 0.091 & V & 19.947 & 0.119 \\
2455701.74844 & 251.6354925 & --45.7130971 & 0.075 & 0.075 & I & 18.875 & 0.048 \\
2455711.50197 & 251.3831321 & --45.7029074 & 0.076 & 0.077 & V & 19.721 & 0.061 \\
2455712.51962 & 251.3562755 & --45.7009425 & 0.078 & 0.079 & I & 18.987 & 0.070 \\
2455712.54178 & 251.3557448 & --45.7009017 & 0.075 & 0.076 & V & 19.735 & 0.052 \\
\multicolumn{1}{c}{\dots} & \multicolumn{1}{c}{\dots} & \multicolumn{1}{c}{\dots} & \multicolumn{1}{c}{\dots} & \multicolumn{1}{c}{\dots} & \multicolumn{1}{c}{\dots} & \multicolumn{1}{c}{\dots} & \multicolumn{1}{c}{\dots} \\
\hline
\end{tabular}
Note: This table is available in its entirety in machine-readable form.
\end{table}

Fluxes were calibrated into standard $I$- and $V$-band magnitudes following standard OGLE procedures \citep{udalski2015}. The typical magnitude precision for deep 2011 observations is 0.045--0.070\,mag, whereas 90\% of shallow observations have photometric precision ranging from 0.087 to 0.290\,mag, with a median of 0.133\,mag.

The positions of the object measured on the subtracted images were transformed into equatorial coordinates using transformations similar to those described by \citet{udalski2015}. These transformations were derived by cross-matching bright stars detected in the OGLE images with the \textit{Gaia} Data Release 3 catalog \citep{gaia2016,gaia_edr3} and calculating the third-degree polynomial transformations between the image coordinates and \textit{Gaia} equatorial coordinates.
Positional uncertainties reported in Table~\ref{tab:results} were calculated by adding in quadrature the centroid fitting uncertainty ($\sigma_{\rm centroid}$) and the root-mean-square (rms) error of the coordinate transformation ($\sigma_{\rm rms}$):
\begin{equation}
\sigma^2 = \sigma^2_{\rm centroid}+\sigma^2_{\rm rms}+\sigma^2_0.
\end{equation}
The value of $\sigma_{\rm centroid}$ ranges from 0.045 to 0.070 pixels for the deep observations and from 0.12 to 0.52 pixels for the shallow observations (with a median of 0.22 pixels). The {\it rms} of the coordinate transformations in one direction was typically $0\zdot\arcs011\!-\!0\zdot\arcs027$, depending on the field. In addition, $\sigma_0=0\zdot\arcs07$ represents a constant error term that was added in order to make sure that $\chi^2/\mathrm{d.o.f.} \approx 1$ for the best-fit astrometric models.

The data are avaiable in the electronic form {\it via}:
\begin{center}
    \textit{https://www.astrouw.edu.pl/ogle/ogle4/2026ED138} \\
    \textit{https://zenodo.org/records/24144451}
\end{center}

\section{Orbital Parameters}

\begin{figure}
\centering
\includegraphics[width=.9\textwidth]{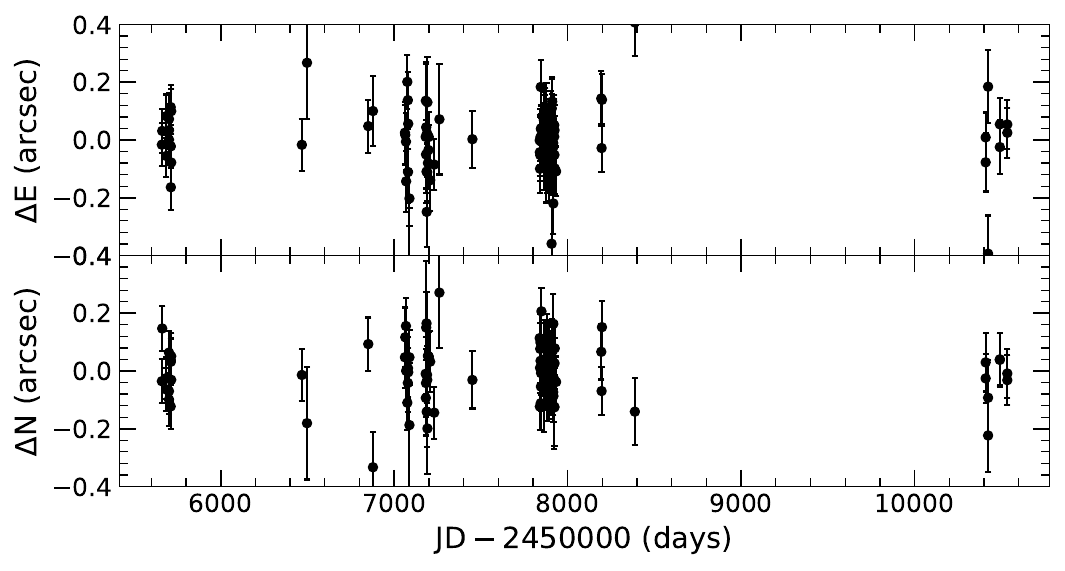}
\caption{Astrometric residuals from the OGLE-only orbital fit.}
\label{fig:residuals}
\end{figure}

We used \textsc{layup} v1.0.0 \citep{holman_zenodo} to derive the orbital elements of 2026~ED138 from the OGLE astrometric dataset. The resulting barycentric ecliptic J2000 orbital elements are presented in the second column of Table~\ref{tab:elements}. The best-fit model yields $\chi^2/{\mathrm{d.o.f.}} = 195.4/198$, indicating a good fit to the data. The astrometric residuals (in the North and East direction) from the best-fit model are shown in Fig.~\ref{fig:residuals}.

\begin{table}[!h]
\caption{Orbital elements of 2026~ED138.}
\label{tab:elements}
\centering
\small
\begin{tabular}{lrrr}
\hline \hline
\multicolumn{1}{c}{Element} & \multicolumn{1}{c}{OGLE only} & \multicolumn{1}{c}{Holman et al.} & \multicolumn{1}{c}{OGLE + Holman et al.}\\
\hline
$a$ (au) & $43.09145 \pm 0.00063$ & $43.09096 \pm 0.00018$ & $43.09105 \pm 0.00016$ \\
$e$ & $0.138737 \pm 0.000007$ & $0.138730 \pm 0.000008$ & $0.138738 \pm 0.000003$ \\
$i$ (deg) & $24.433904 \pm 0.000019$ & $24.433893 \pm 0.000009$ & $24.433894 \pm 0.000006$ \\
$\Omega$ (deg) & $322.60583 \pm 0.00009$ & $322.60591 \pm 0.00003$ & $322.60591 \pm 0.00002$ \\
$\omega$ (deg) & $168.70672 \pm 0.00448$ & $168.70034 \pm 0.00284$ & $168.70335 \pm 0.00128$ \\
$M$ (deg) & $124.85482 \pm 0.00518$ & $127.58242 \pm 0.00402$ & $127.57815 \pm 0.00159$\\
\hline
$\chi^2/\mathrm{d.o.f.}$ & $195.4/198$ & $83.3/142$ & $284.9/346$ \\
Epoch & 60423.08984 & 61203.02256 & 61203.02256 \\
(MJD TDB) & & & \\
\hline
\end{tabular}
\end{table}

In order to validate the accuracy of the OGLE astrometric measurements, we used \textsc{layup} to fit the orbit to the astrometric measurements reported by \citet[][Table~11]{holman2026}, which are based on DECam and Rubin observations collected from 2016 through 2026. This fit yielded $\chi^2/\mathrm{d.o.f.}=83.3/142$. The best-fit parameters are reported in the third column of Table~\ref{tab:elements} and they match the parameters from the OGLE-only fit within their uncertainties (except for the mean anomaly $M$, which was calculated for a different epoch). 

\begin{figure}
\centering
\includegraphics[width=.9\textwidth]{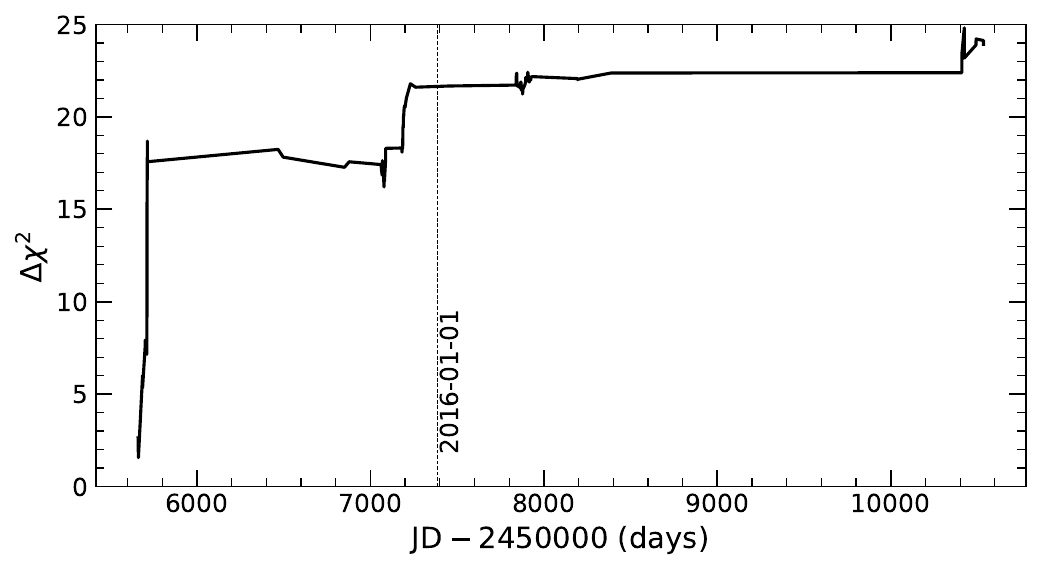}
\caption{Cumulative $\chi^2$ difference between the astrometric model that was fit to the \citet{holman2026} data and that fit to the OGLE-only data.}
\label{fig:delta_chi2}
\end{figure}

We then used the DECam+Rubin model to calculate the predicted positions of 2026~ED138 at the time of OGLE observations. Evaluating the OGLE observations against this DECam+Rubin orbit yielded $\chi^2 = 219.3$, which was $\Delta\chi^2=23.9$ larger than for the OGLE-only fit. This difference can be mostly explained by the fact that the OGLE dataset includes 41 pre-2016 observations, which significantly extend the observational arc. These 41 epochs contribute to $\Delta\chi^2=21.6$, while the remaining $\Delta\chi^2$ difference can be attributed to the statistical noise in the data. Fig.~\ref{fig:delta_chi2} shows the cumulative $\chi^2$ difference between the astrometric model that was fit to the \citet{holman2026} data and that fit to the OGLE-only data.

The fourth column of Table~\ref{tab:elements} presents the orbital solution derived from the combined OGLE, DECam, and Rubin datasets.

Finally, we compared our astrometric observations with the orbital parameters calculated by \citet[][Tables~4 and 5]{murtagh2026}. We used their model to calculate the predicted positions of 2026~ED138 at the time of OGLE observations. For that model, we found $\chi^2=203.5$, i.e., a $\chi^2$ difference of $\Delta\chi^2=8.1$ compared to the OGLE-only model, indicating a slight tension. This $\Delta\chi^2$ difference cannot be attributed to data from a specific night or field, but rather accumulates throughout the entire duration of the observations. 

\section{Phase Curve and Search for Periodic Variability}

\begin{figure}
\centering
\includegraphics[width=.8\textwidth]{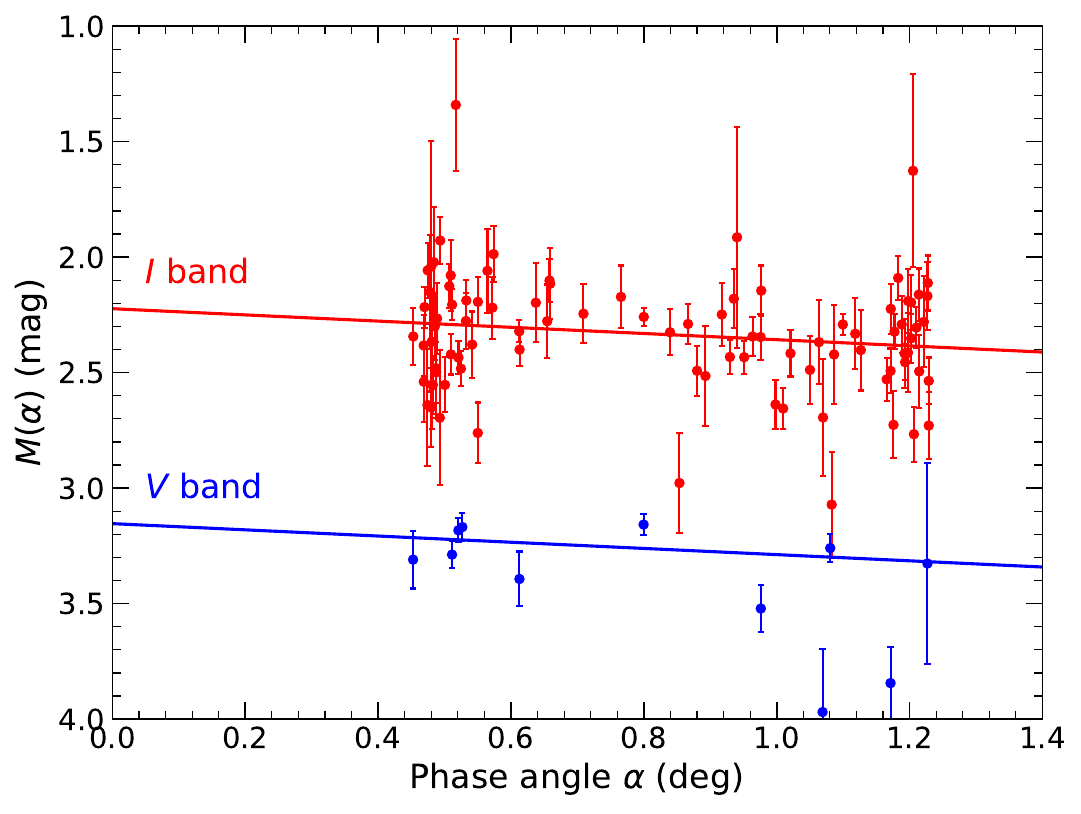}
\caption{OGLE phase curve of 2026~ED138. Red and blue lines mark the best-fit models for the $I$- and $V$-band observations, respectively.}
\label{fig:phase}
\end{figure}

The OGLE dataset spans over 13 years of observations across a relatively broad range of phase angles, ranging from 0.45 to 1.23\,deg. Moreover, because all observations were obtained using standard $I$- and $V$-band filters, we can directly determine the absolute magnitudes of 2026~ED138. To this end, we fitted the following model to the OGLE $I$- and $V$-band data:
\begin{equation}
m_\lambda = H_{\lambda} + 5 \log_{10}(r\Delta) + \beta\alpha,
\end{equation}
where $r$ and $\Delta$ are the heliocentric and geocentric distances, $\alpha$ is the Sun--target--observer phase angle, $\beta$ is the phase slope (common to both bands), and $H_\lambda$ is the absolute magnitude in a given band. Fig.~\ref{fig:phase} shows the dependence of the reduced magnitude, $M_\lambda(\alpha) = m_\lambda - 5 \log_{10}(r\Delta)$, on the phase angle ($\alpha$).

A linear least-squares fit yields:
\begin{align}
\begin{split}
H_I &= 2.224 \pm 0.034\,\mathrm{mag}, \\
H_V &= 3.154 \pm 0.035\,\mathrm{mag}, \\
\beta &= 0.134 \pm 0.038\,\mathrm{mag}\,\mathrm{deg}^{-1}.
\end{split}
\end{align}

These values are consistent with those reported by \citet{holman2026} ($H_V = 3.19 \pm 0.04$\,mag and $\beta=0.14 \pm 0.03$\,mag\,deg$^{-1}$) and \citet{murtagh2026} ($H_V = 3.11 \pm 0.05$\,mag and $\beta = 0.14 \pm 0.04$\,mag\,deg$^{-1}$). Unlike those studies, which derived $H_V$ indirectly from $g$- and $r$-band observations, our results are based on direct $V$-band photometry.

We derived a $V-I$ color index of the object to be $V-I = H_V - H_I = 0.930 \pm 0.026$\,mag. (Note that the uncertainty on $V-I$ is smaller than the uncertainty on either $H_V$ or $H_I$ because these values are strongly correlated.) Comparing this color index with the average values for taxonomic classes from \citet{barucci2005}, we place 2026~ED138 in the BR group (intermediate blue-red), consistent with the classification of \citet{murtagh2026}.

\begin{figure}
\includegraphics[width=\textwidth]{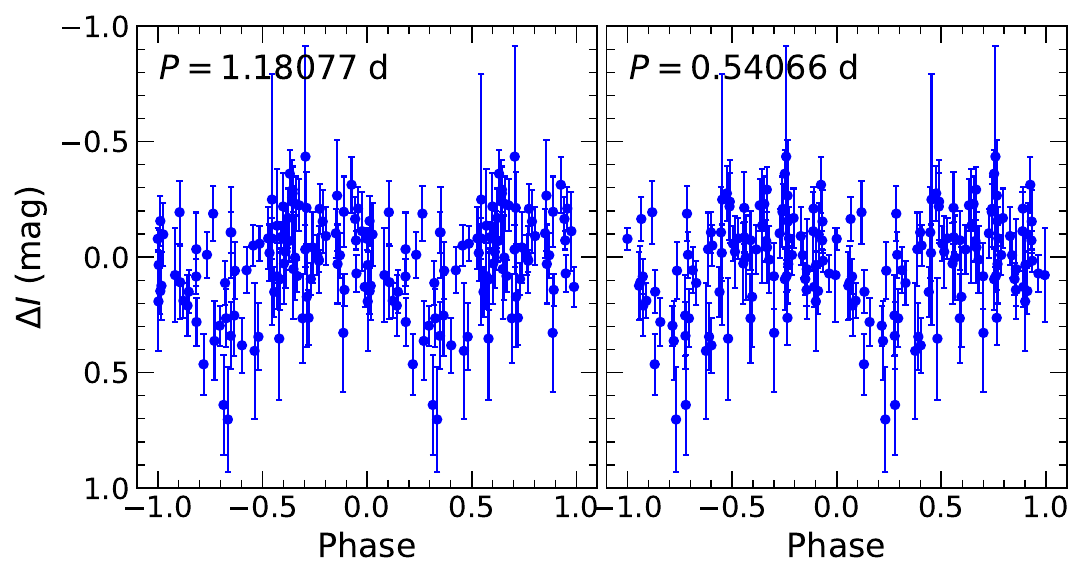}
\caption{OGLE phase-angle-corrected $I$-band light curve of 2026~ED138 folded with possible periods of $P=1.18077$\,d (left panel) and $P=0.54066$\,d (right panel).}
\label{fig:phased}
\end{figure}

The phase-angle-corrected OGLE photometry of 2026~ED138 was subjected to
period-finding procedures. We searched for possible stable periods that
could be interpreted as rotation periods of the dwarf planet
candidate. We used two period-search algorithms: {\sc FNPEAKS}\footnote{\url{http://helas.astro.uni.wroc.pl/deliverables.php?lang=en&active=fnpeaks}} and
{\sc TATRY} \citep{schwarzenberg1996}. Both searches yielded a similar
peak frequency at $P=1.18077$\,d. However, the S/N of the frequency peak is at a relatively low level ($\mathrm{S/N}=3.6$), making this detection very uncertain. Moreover, when two significant outliers were removed from the light curve, an alternative periodicity of $P=0.54066$\,d showed up with similar strength. This is
likely a 1-day alias of the former period. Nevertheless, we are unable
to definitively conclude if these periodicities are real, and both must
be treated as very tentative. The OGLE $I$-band data phased with these two periods are presented in Fig.~\ref{fig:phased}. 

Note that \citet{murtagh2026} reported tentative detection of periodic variability with a period of 0.4317\,d at $\mathrm{S/N}=2.2$ based on the analysis of DECam, Rubin, SkyMapper, VISTA, and ATLAS data. We do not find any significant signal at similar period in the OGLE data. Continuous observations of 2026~ED138 conducted over several consecutive nights will be required to confirm whether coherent rotational variability is present.

\section{Conclusions}

The discovery of a TNO as bright and potentially large as 2026~ED138 by \citet{holman2026} was unexpected. Wide-field sky surveys conducted over the past two decades, including the OGLE OCKS, covered and searched the sky regions most likely to harbor large TNOs, including dense fields in the Galactic bulge. However, this discovery demonstrates that orbital inclinations of TNOs may be higher than previously expected, suggesting that some such large objects may still remain undiscovered.

Although 2026~ED138 evaded detection by the OGLE OCKS in June 2010 because it was located approximately 3\,deg south of the survey's ecliptic latitude limit, it crossed several OGLE fields over the subsequent 16 years. The 102 OGLE pre-discovery detections of this dwarf planet candidate (including 41 epochs before 2016) almost doubled the available observational dataset. Extending the observational arc back to 2011 using these high-precision CCD measurements significantly refines the orbital solution. Furthermore, OGLE standard $VI$ photometry led to the direct determination of the planet's $V$-band absolute magnitude of $H_V=3.154 \pm 0.024$\,mag, and the ${V-I}$ color index equal to $0.930\pm 0.026$\,mag, as well as a tentative rotation period of 1.18077\,d or 0.54066\,d.

The high intrinsic brightness of 2026~ED138 suggests that this dwarf-planet candidate may have a large size, with an estimated diameter ranging from 612 to 1530\,km \citep{holman2026} depending on its unknown albedo. Over the coming years, the object will continue to drift across dense stellar fields south of the Galactic bulge. Thus, the chances of stellar occultations, which may lead to a direct size estimate, are relatively large. 

\citet{murtagh2026} provided a list of such phenomena in the coming years based on their orbital parameters. As these predictions are extremely sensitive to the accuracy of the orbit, they can be verified with our determination of orbital parameters of 2026~ED138. This is, however, out of the scope of this paper.

\section*{Acknowledgments}

The OGLE project has received funding from the Polish National Science Centre grant OPUS 2024/55/B/ST9/00447 awarded to A.U.

\bibliographystyle{acta}
\bibliography{pap}

\end{document}